# Narrowband wide dynamic range tunable filter based on Fano resonant planar multilayered structure

Ibrahim Abdulhalim

Department of Electrooptic and Photonics Engineering and the Ilse-Katz Center for Nanoscale Science and Technology, Ben Gurion University, Beer Sheva 84105, Israel

abdulhlm@bgu.ac.il

Abstract:

Tunable narrowband spectral filters with high light throughput and wide dynamic range have remarkable applications such as in optical communications, optical spectroscopy and spectral imaging. However, a cost is usually associated with the filter narrowing either in the dynamic range, in the throughput or the manufacturability. Here we report on a resonating planar multilayered structure that exhibits transparency window in reflection with a controllable full width at half maximum (sub-Angstroms till tens of nm) and tunability over wide spectral range (>500nm in the visible and near infrared). The phenomenon is observed in TE and TM polarizations with much higher contrast in TE. Fano type resonance originating from coupling between waveguide modes and lossy surface electromagnetic waves supported by field distribution calculations explains the phenomenon. The wide tuning range with high contrast is mainly achieved using an absorptive layer with high imaginary to real part ratio of the dielectric constant that enables excitation of lossy surface waves known to exist over a wide spectral band in thin films. To avoid large losses, it is found that the lossy layer should be ultrathin (6nm Cr layer for example). The tuning is achieved by small angular scan of less than 2 degrees or by modulating the refractive index or thickness of the submicron thick waveguide layer from the visible till the near infrared range and in principle it can be designed to operate in any spectral range. Such a thin variable index or thickness layer can allow tuning at ultrahigh speed using conventional electrooptic, magnetooptic, piezoelectric or thermooptic materials at relatively low external fields.

Resonant optical structures reveal transmission, reflection or absorption peaks at specific wavelengths or angles, which are highly sensitive to the surrounding media or to the parameters of the structure itself. Examples include Fabry-Perot cavities, resonant gratings, grating coupled waveguide structures, whispering gallery modes, surface plasmon resonances, Bloch surface waves, Tamm surface waves and more [1-3]. Coupling between two or more resonators is another emerging methodology for obtaining high transparency window through what is called Fano resonances, electromagnetically induced transparency or plasmon induced transparency [3]. With the rising metamaterials field and the progress in nanofabrication of different structures in large areas this whole topic of resonance structures design and manufacturability is becoming increasingly hot due to its technological importance as well as the ability it provides to demonstrate fundamental physical phenomena [4]. The phenomenon of the appearance of almost perfect transparency in a narrow spectral window in an otherwise absorbing system is a phenomenon that appears in several fields such as atomic physics, plasmonic materials and even classical systems of coupled oscillators [4]. Together with the intriguing phenomenon there appears the slowing down of light and even possible stopping of light pulses. However, the narrow resonance which might act as an optical filter is usually limited in its dynamic range and its manufacturability requires very precise materials parameters and alignment processes which increase the cost and usually limit its size. In electromagnetic induced transparency, since the absorption band is already narrow, the transparency window tunability range is limited [4]. Even in non-absorbing resonant structures such as the guided mode resonance explained as Fano type resonance, the tuning range is limited to few tens of nm up until a maximum of 100nm for optimized structures [6]. In an attempt to increase the tunability, I have searched for materials and surface waves which exhibit high absorption in a wide spectral range such as lossy surface electromagnetic waves (LSEWs) or sometimes called lossy mode resonances (LMRs) [7-10]. These waves have been shown to exist in absorbing thin films and they are more pronounced when the imaginary part of the dielectric constant is much larger than the real part, for example to demonstrate perfect absorption in ultrathin layer [11] and single layers of graphene [12,13]. Here I propose a planar structure that exhibits controllable bandwidth resonances over a wide dynamic range covering over 500nm in the visible and the near infrared ranges. Its tunability can be achieved by tuning a sub-micron electrooptic, magnetooptic, thermooptic or piezoelectric layer with relatively small external fields or by simple angular scanning. Hence the proposed structure has a great potential for spectral imaging, Raman spectroscopy, infrared spectroscopy, sensing and the optical telecommunications industry. Such a narrow tunable filter can be used in many fields such as telescopes, to discover life in space or in miniature biosensors, to detect pollutants in the environment, food, agriculture, water quality, blood analytes, or with an endoscope inside the human body to diagnose diseases or to boost the capacity and speed of the telecommunication channels using the wavelength division multiplexing approach. Using fast modulation materials, it can also be used for generating intense laser pulses through Q-switching or mode locking of lasers.

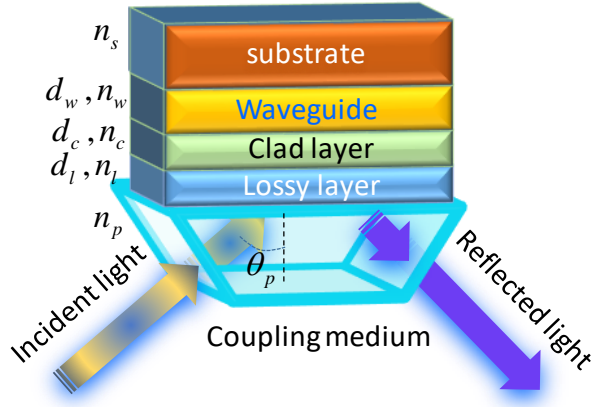

Figure 1: Schematic drawing of the resonating planar multilayered structure.

The geometry of the structure is shown in figure 1. It consists of a three-layer planar structure deposited on a prism. The first layer is the ultrathin lossy layer $l$ necessary to generate the lossy surface waves. This layer does not need to be metallic as the only requirement for lossy wave in thin films is to have non-zero, and preferably large imaginary part of the dielectric constant $\varepsilon_i$ without condition on the negative real part $\varepsilon_r$. When the material has negative real part it becomes plasmonic, although we shall show that even with metals having $\varepsilon_i \gg |\varepsilon_r|$ such as Cr, it is possible to excite the LSEWs in ultrathin films down to 6nm thickness. The optimum thickness for maximum absorption in the prism coupling geometry can be easily estimated for TE wave using the Airy formulae for reflection and transmission of a single layer, writing the absorption as: $A = 1 - R - T$ and finding the conditions for maximum absorption. Assuming the absorptive layer on the prism under total internal condition and ignoring $\varepsilon_r$, the final result for the optimum thickness of the lossy layer is:

$$d_{l-op} = \frac{\lambda \sqrt{n_p^2 - 1}}{2\pi \varepsilon_i} \quad (1)$$

For Cr, in the visible range we have $\varepsilon_i \approx 20 - 22$, giving $d_{l-op} \approx 5 - 7nm$ which agrees with the optimum thickness found to give the best contrast as will be seen below. The 2nd layer acts as a clad, semi-clad or coupling layer $c$ to the 3rd waveguide layer $w$. The substrate $s$ is a low index semi-infinite substrate, which for biosensing application can be the environment medium such as water or air via the evanescent wave in the substrate. The existence of the lossy layer is crucial to generate the lossy surface electromagnetic wave (LSEW) which can then allow the excitation of the guided mode (GM). Its thickness preferably be very small in order to minimize the losses as we are seeking a high reflectance peak, and here we will present a design in which 6nm of Cr is an optimum thickness for the visible and near infrared ranges. The thickness of the clad or semi-clad layer controls the coupling between the LSEW and the waveguide, hence the width of any resonance that may emerge is expected to be controlled by modulating this layer. When it is very thick, the escape of the radiation from the waveguide becomes more difficult and the resonance will be satisfied only at very narrow range. At moderate thicknesses comparable to the wavelength or less, more radiation can escape the waveguide at the resonance. Hence one expects to control the width of the resonance by varying the optical thickness of this layer, which is why we refer to it also as the coupling layer.

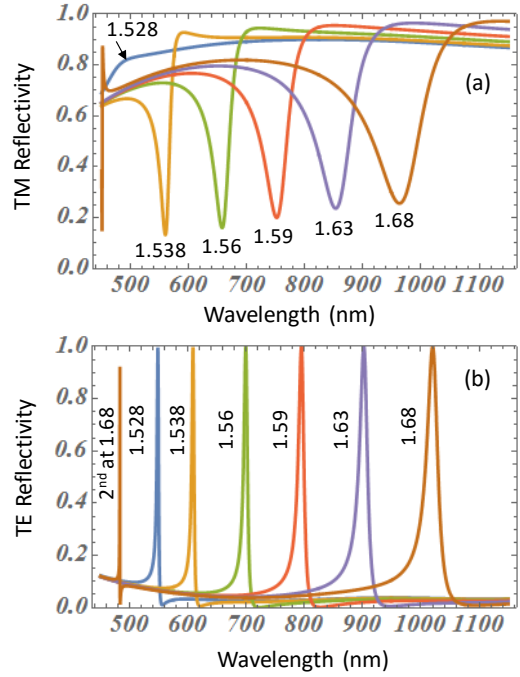

Figure 2: Calculated reflectivity spectra from the stack of figure 1 using the following layers: 6nm Cr/575nm MgF2/603nm variable index layer/MgF2 substrate. The incidence angle in the SF11 prism is $\theta_p = 52°$. (a) TM, (b) TE.

To simulate the reflection properties of this structure the 2x2 transfer matrix method of Abeles is used. Figure 2 shows results for TM (a) and (TE) reflectivity spectra. The TM reflection spectrum shows dips because of the lossy wave excitation, while the TE reflection spectrum shows narrow peaks, both are tuned by varying the refractive index of the waveguide layer over a range of 500nm. Then 2$^{nd}$ order peak for the 1.68 case corresponding to the coupling between the lossy mode and next waveguide mode TE1.

Note that the refractive indices of the variable index layer are comparable to those of liquid crystals and the refractive index change required is nearly 0.152 which is easily achievable with ordinary nematic liquid crystal such as E7 or 5CB. Obtaining a narrow peak over such large tuning range is superior over many resonant structures and it will be shown later that the peak width can be made narrower by changing the optical thickness of the clad layer. As a comparison to other resonant devices, to achieve 500nm free spectral range (*FSR*) of a Fabry-Perot cavity the thickness of the cavity should be around $d = \lambda^2 / 2nFSR \approx 350nm$ by taking the value for wavelength of 750nm near the middle of the range in figure 2b. In this case, the finesse required to get to full width at half maximum of FWHM=2nm will be $F = FSR/FWHM \approx 175$ which imposes high reflectivity condition on the cavity mirrors of more than 98%, making it more difficult to manufacture as well as requiring liquid crystal birefringence of $\Delta n = 2d / FSR \approx 0.71nm$, which is very high and difficult to achieve. Hence the proposed tunable filter is superior over cavity and other resonant tunable structures. To vary the FWHM of the filter it is enough to increase the thickness of the cladding layer as shown in figure 3. Note that the dip in TM and the peak in TE shift towards higher wavelengths as the clad layer thickness increases. This is due to the increase of the phase accumulated in this layer as well as because the effective index of the GM is affected by the thickness of the clad layer when it is comparable to the wavelength.

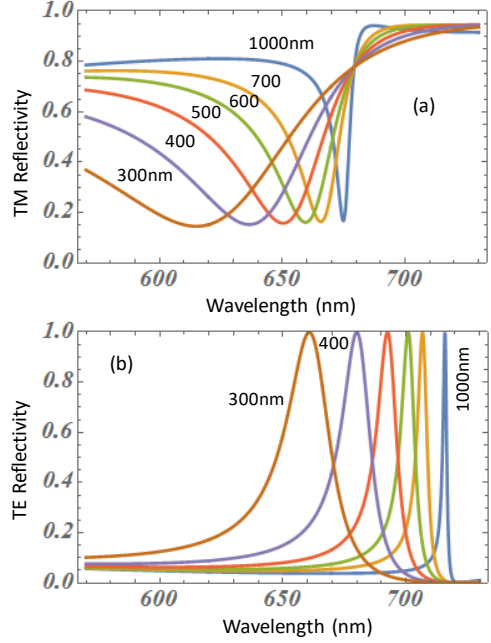

Figure 3: Calculated reflectivity spectra for the same structure of figure 2 at fixed refractive index of the waveguide layer of 1.56 but variable thickness of the clad layer. (a) TM, (b) TE.

To explain the shift and the dependence on the cladding layer thickness let us write the phase difference between the transmitted and reflected waves which may be written as:

$$\varphi_{t-r} = \varphi_w + \varphi_{TIR} + \varphi_c + 2\varphi_{cou} \qquad (2)$$

Here $\varphi_w = 2k_w d_w$, $\varphi_c = 2k_c d_c$ are the phases accumulated during a round trip in the waveguide and cladding layers, $\varphi_{TIR}$ is the phase due to total internal reflection at the waveguide interfaces and $\varphi_{cou} = 2\varphi_{Fresnel} + 2\delta$ is the phase due to the Fresnel phases at the lossy layer interfaces and the additional phase $\delta$ is due to the lossy wave excitation or coupling from the prism/lossy layer system and back to the prism. The guided wave condition is given by the well-known phase matching condition:

$$2k_w d_w + \varphi_{TIR} + 2\varphi_{Fresnel} = 2\pi n \qquad (3)$$

Here *m* being an integer, revealing to:

$$\varphi_{t-r} = 2\pi n + 2\delta + 2k_c d_c \qquad (4)$$

For a constructive interference to occur in reflection the condition is:

$$\varphi_{t-r} = 2\pi m + 2\delta + 2k_c d_c = (2j+1)\pi \tag{5}$$

Here j is another integer and designating $q = j - m$, we can write:

$$2k_c d_c = (2q+1)\pi - 2\delta \tag{6}$$

As $d_c$ increases the wavelength has to increase in order to maintain this condition satisfied because the wave vector $k_c$ is inversely proportional to the wavelength. Similarly from equation 3, the peak or dip are red shifted with the change of the waveguide layer index or thickness because $k_w d_w$ is proportional to the refractive index and thickness while inversely proportional to the wavelength. The changes of the FWHM as $d_c$ changes may be understood from the fact that when the cladding layer is too thin then the mode will leak more from the waveguide causing broadening of the reflected peak. In this sense the semi-clad layer maybe considered as a pump to transfer the energy from the lossy mode to the waveguide and from the waveguide to the free space. This energy transfer from the waveguide maybe expressed by a loss factor or more accurately "leak factor" $\gamma_{leak}$ which controls the quality factor of the resonance: $Q = \omega_0 / \gamma_{leak}$ with $\omega_0$ being the resonance frequency. It also determines the width of the resonance.

The fact that the reflectivity in TM consists of a dip-only without any signs of a peak indicates that the TM wave is not interacting efficiently with the waveguide or weakly interacting. This will be verified later based on the field distribution calculations. As one expects, the cutoff condition for TE guided mode excitation is smaller than that for TM. One can increase the interaction of the TM LSEW with the waveguide by decreasing the wavelength or increasing the angle. In fact, figure 2a shows some signs of a weak peak at the shortest wavelength spectrum around 500nm superimposed to the dip. To enhance this further we performed simulations for the same structure of figure 1 but at larger angle of incidence $\theta_p = 55°$ (see figure 4). In figure 4a the TM dip becomes clearly similar to Fano type resonance shape indicating the coupling between two resonators. This is expected in TM waves similar to the case of plasmon coupled waveguide modes observed recently [14]. To strengthen further the fact that the emerging transparency windows are due to coupling between lossy surface waves and waveguide mode we calculated for variety of materials shown in figure 6 including metals.

Note that when Ag, Cu are used an SP wave is excited in TM polarization and the transparency window appears with high contrast at the center of the SPR dip. This is well known now and investigated by several authors both using extended plasmons geometry [14] as well as other plasmonic and metamaterial configurations [15-18], but we would like to highlight the fact that Cu and Al are not suitable plasmonic materials for the near infrared range showing very broad peak, wherein here our original thought was to use this fact in order to increase the tunability range of the transparency window. Besides this, note that for VO$_2$ at 20 °C, which is an insulator, only lossy mode is excited with low contrast while VO$_2$ at the metallic phase (100 °C) shows beginning of very broad minimum with superimposed peak, although it becomes plasmonic ($\varepsilon_r < 0$) only above 1200nm wavelengths. The reason for the broad minimum is due its larger ratio of $|\varepsilon_i/\varepsilon_r|$. Note that for Al, only 7nm is required, while for thin Al layer on a prism there is no symmetric SPR dip in the visible range and usually the minimum reflectance is not close to zero, while here we get near zero dip and high contrast peak, hence we conclude that even for Al in this case the wave is a lossy mode.

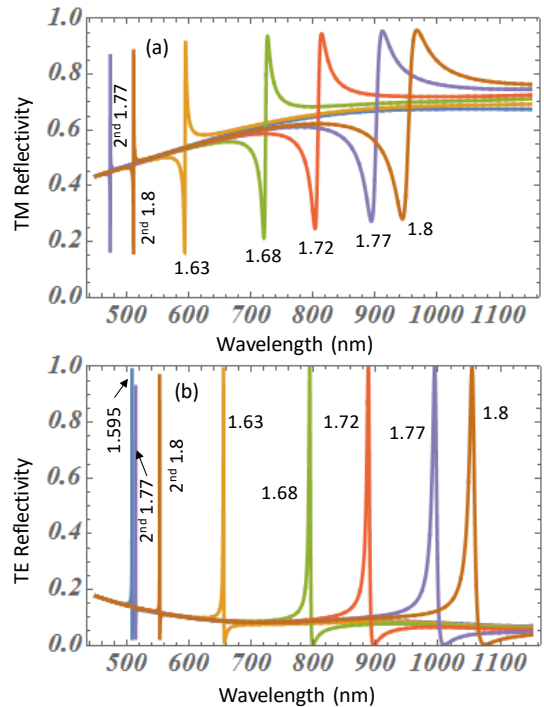

Figure 4: TM (a) and TE (b) spectra calculated for the same structure of figure 1 at an angle of $\theta_p = 55°$ to increase the interaction of the TM LSEW with the waveguide.

For TE polarization VO$_2$ is showing pronounced peaks, although with low contrast for the non-metallic phase at 20 °C, it demonstrates the possibility of modulating the peak thermooptically as this material is well known to have strong thermooptic property and undergoes insulator to metal transition at around 68 °C. Although being metallic, the real part of the dielectric constant becomes negative only at the infrared wavelength starting from nearly 1200nm, which is why it is used as smart window [19]. The better contrast obtained for the metallic phase in figure 5b is because the ratio $|\varepsilon_i/\varepsilon_r|$ is larger (~4) in the metallic phase compared to the insulating phase (~1/3). At the relatively high thicknesses of Ag and Cu, the structure is mainly reflecting at TE polarization and no coupling to the waveguide occurs. The peaks can also be observed at specific angles, which is important to show as the modulation at a single wavelength is very important for generating laser pulses and precision measurement or sensing applications. Figure 6 shows calculated reflectance versus angle inside the prism medium at different thicknesses of the waveguide layer made of AlN as this material is known to be piezoelectric. Almost 100% modulation can be reached with few nm modulation of the thickness at the optical telecommunication window wavelength of 1550nm.

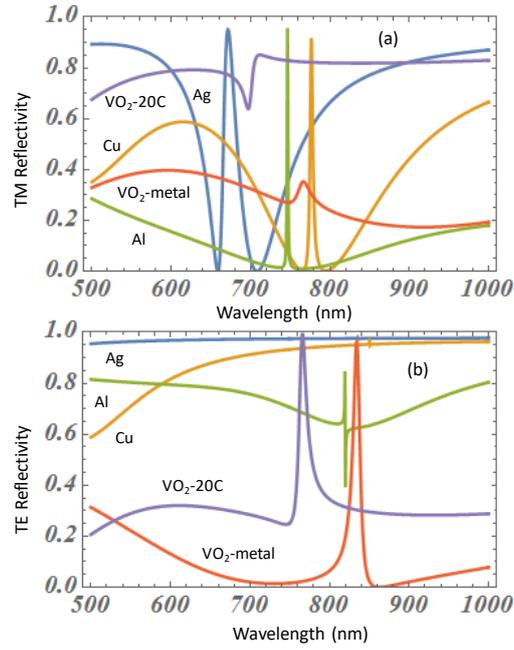

Figure 5: Calculated reflectance spectra for variety of materials including metals and non-metals to demonstrate the generality of the observed phenomenon to lossy materials. Parameters of the calculations for each material are given in table 1. (a) TM, (b) TE.

Table 1: Parameters of the layers used in the calculations of figure 5. The clad layer is MgF$_2$ while the lossy material is variable and the refractive indices and thicknesses are optimized to give the optimum contrast.

| Material | $\theta_p$ (°) | $d_l$ (nm) | $d_c$ (nm) | $d_w$ (nm) | $n_w$ |
|---|---|---|---|---|---|
| Ag | 55.2 | 46 | 700 | 196 | TiO$_2$ |
| Cu | 55 | 40 | 1000 | 700 | 1.65 |
| Al | 55.5 | 7 | 800 | 700 | 1.652 |
| VO$_2$-metal | 55.5 | 45 | 400 | 700 | 1.66 |
| VO$_2$-insulator | 55.5 | 20 | 400 | 650 | 1.66 |

Another interesting property highlighted in figure 6c and that is the large angular sensitivity of the resonance wavelength showing that one can scan nearly 500nm over 2.5 degrees. The red shift with the incidence angle can be explained by the fact that one expects momentum matching at the resonance, meaning the tangential component of the incident wave vector $k_x = 2\pi n_p \sin\theta_p / \lambda$ has to be conserved. Hence increasing the angle must be accompanied by an increase in the resonance wavelength. As a refractive index sensor the refractive index of the substrate should be variable and in figure 6d the shift of the resonance peaks with the analyte index is shown for refractive indices from 1 till 1.38. Note that the sensitivity is higher in the high index side reaching around 800nm/RIU while it drops to around 200nm/RIU in the low index side. This is because the wave becomes more confined in the waveguide layer as the substrate index decreases. However, the peak becomes narrower by the same factor, meaning the resolution will not change.

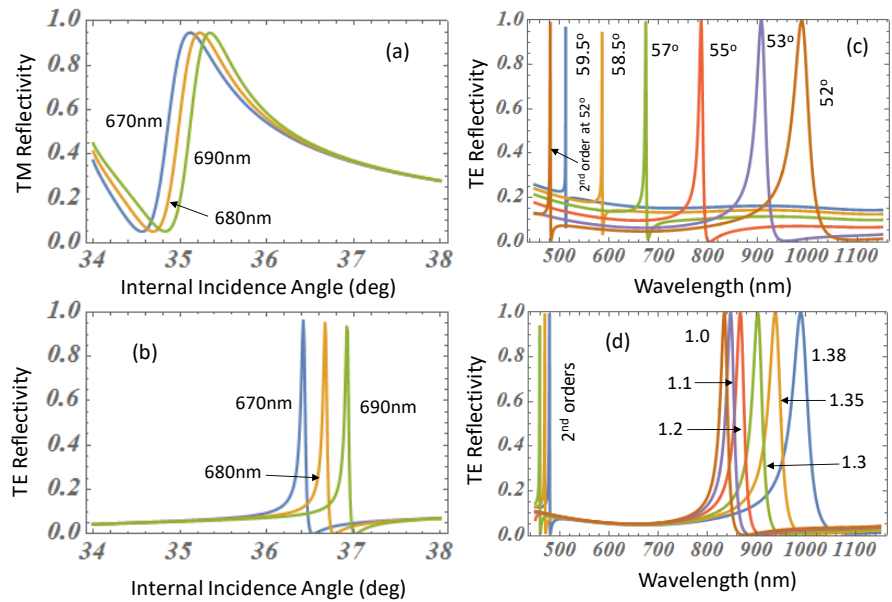

Figure 6: (a) and (b): Reflectance versus angle in the prism medium at different thicknesses of the AlN waveguide layer. The structure is: TiO$_2$ prism/15nm Cr/800nm MgF$_2$/AlN with variable thickness/MgF$_2$ substrate. (a) TM, (b) TE. (c) and (d) show the TE reflectivity spectra at different incidence angles (c) or different substrate refractive indices (d). In (c) and (d) the prism is SF11, 6nm Cr, 400nm MgF$_2$, 603 waveguide layer at refractive index 1.68. In (d) the incidence angle is 52 degrees while in (c) the substrate is MgF$_2$.

To verify that the waveguide mode is indeed excited at the resonance, we have used both the analytic approach based on the 2x2 Abeles matrices as well as COMSOL multiphysics to simulate the field distribution. Figure 7 shows the intensity distribution of the y-component of the electric field. From left to right, at 730nm far away from the resonance corresponding to refractive index 1.59 of figure 2b (resonance at 794.5nm), the field is mainly in the prism medium, that is no interaction between the lossy mode and the waveguide mode. The reflection at 730nm is only 5% according to figure 2b and since there is no transmission (incidence angle above the critical angle), the conclusion is that most of the energy at this wavelength and all wavelengths around the peaks get absorbed. This is the essence of the broadband lossy mode and our proposition to use this fact for wide tuning range. As the wavelength becomes closer to the resonance wavelength the energy becomes more confined in the waveguide and at the resonance it is totally in the waveguide. Hence this is the proof that the resonance corresponds to the guided mode, but due to the interaction with the lossy mode through the thin clad layer it escapes the waveguide in a Fano type resonance shape. To verify that the TM is not a waveguide mode we calculated the field distribution for the TM wave around the corresponding dip in figure 2a at 752nm as presented in figure 8. It clearly shows that most of the energy is outside the waveguide, hence the mode is mainly the lossy mode which gives dip corresponding to absorption in the lossy layer.

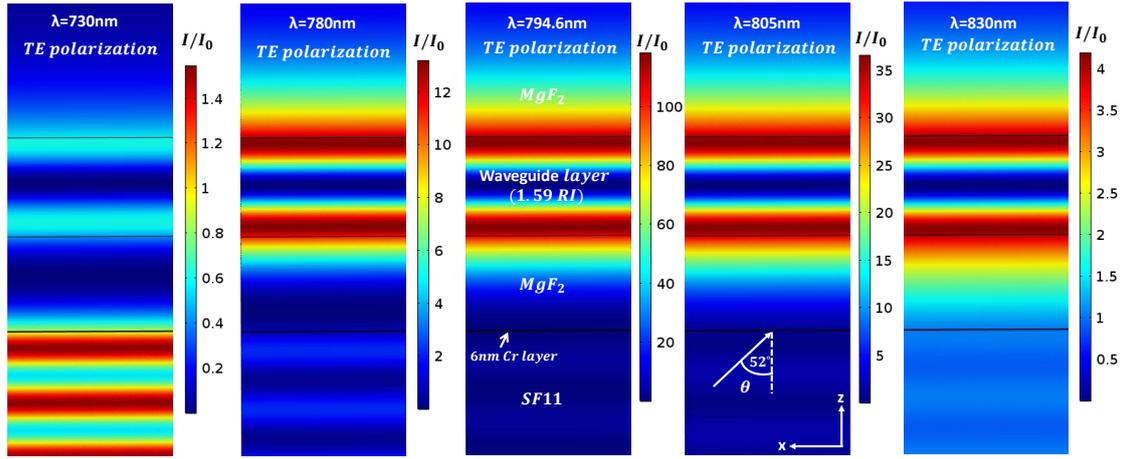

Figure 7: TE polarization electric field intensity distribution $|E_y|^2$ for the structure of figure 2b at different wavelengths around and at the resonance of 794.6nm.

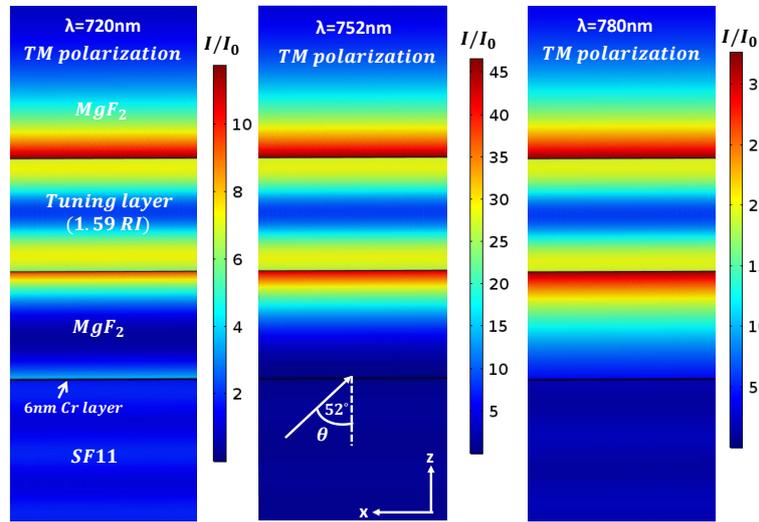

Figure 8: TM polarization electric field intensity distribution $|E_{tot}|^2 = |E_x|^2 + |E_z|^2$ for the structure of figure 2a at different wavelengths around and at the dip of 752nm.

To understand the analogy between the observed phenomenon, the Fano resonance and the coupled classical oscillators, in figure 9 we present a diagram for illustration. In figure 9a a cross section side view of the structure is shown with the LSEW generated at the lossy layer-clad interface and the GM wave is in the waveguide. The quantum mechanical analog is drawn with the zero level representing the incident light from the continuum (prism medium), the 1st energy level with some width determined by the damping $\gamma_1$ represents the frequency of the LSEW ($|1\rangle \equiv |LSEW\rangle$) while the 2nd level represents the GM frequency ($|2\rangle \equiv |GM\rangle$) with damping $\gamma_2$. According to Fano [20] there are two possible transition paths that interfere together to give the well-known Fano shape resonance: the direct transition $|0\rangle \rightarrow |1\rangle \equiv |LSEW\rangle$ and the indirect one $|0\rangle \rightarrow |1\rangle \equiv |LSEW\rangle \rightarrow |2\rangle \equiv |GM\rangle \rightarrow |1\rangle \equiv |LSEW\rangle$. The classical analog is depicted in

figure 9c which shows two coupled oscillators having coupling constant $\kappa$ provided by the clad layer which is acting as a coupling medium. Without the clad layer, using such an ultrathin lossy layer the coupling to the GM will be over broadband and weak but as the clad layer thickness increases the coupling increases and becomes very strong at the resonances with increasing quality factor as it was shown in figure 3. The damping constant $\gamma_2$ is very small as the waveguide material is transparent which fits the requirement that one should have $\gamma_2 \ll \gamma_1$. The normalized shape of Fano resonance can be expressed by [20]:

$$R_F = R_b + \frac{1}{1+S^2} \frac{(2(\omega-\omega_0)/\Delta + S)^2}{4(\omega-\omega_0)^2/\Delta^2 + 1} \qquad (7)$$

Here $R_b$ is the background reflectivity, for example from the prism-lossy layer interface, $\Delta$ is the FWHM and $S = S_0 \sin\phi$ is a shape parameter which can have positive and negative values with the phase term $\sin\phi$ being introduced to take care of the sign. The phase $\phi$ maybe considered as a result of the phase difference between the two oscillators. In figure 9e and 9d different Fano shapes are drawn using FWHM of 6nm, $R_b = 0$, and different shape parameters. At S=0 a pure dip is observed while as S becomes very large in its absolute value a pure peak is obtained and if we compare to our results in figure 2b they correspond to large S around 8-10 while the TM case corresponds to small S values. Hence we may conclude that to get a peak with large contrast, a high S value is needed. For intermediate S values the shape has a peak superimposed on a dip. The interference between the two transition paths is actually expressed when S has intermediate values. To explain this, we consider equation (7) and writing it in the following form:

$$R_F = R_b + \frac{1}{(1+S^2)(4(\omega-\omega_0)^2/\Delta^2 + 1)} \left[4(\omega-\omega_0)^2/\Delta^2 + S^2 + 4(\omega-\omega_0)S/\Delta\right] \qquad (8)$$

The 1st and 2nd terms in the squared brackets give a peak and a dip shapes at the resonance respectively, while the 3rd term is the interference term which depends on the sign of S determined by the phase difference $\phi$. This 3rd term vanishes at the resonance but it affects the symmetry of the Fano resonance depending on the value and sign (phase) of S.

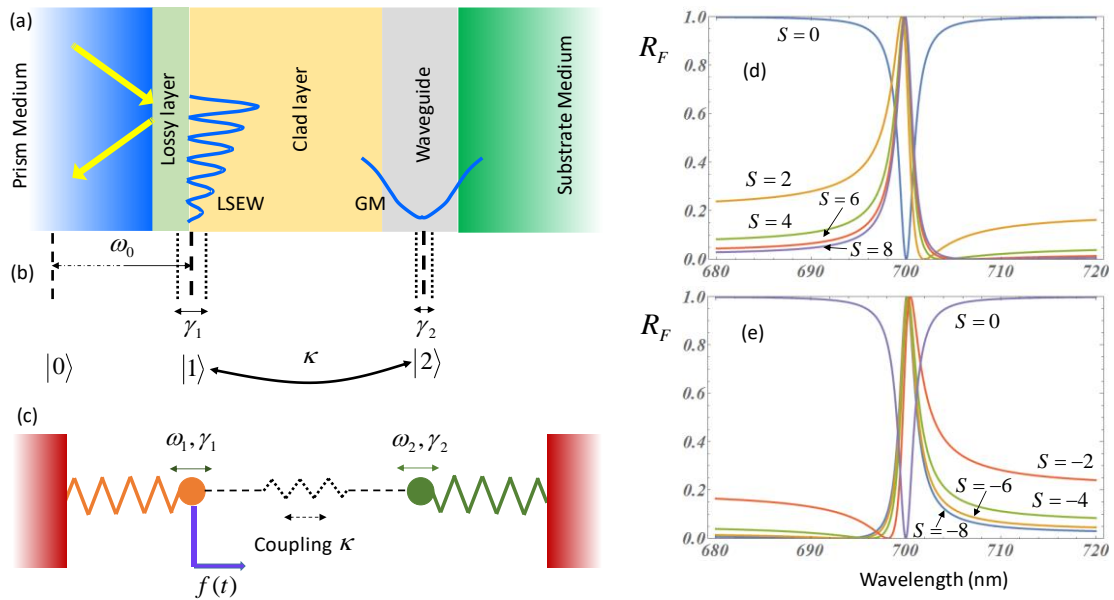

Figure 9: Schematic representation of the structure giving the two interacting waves LSEW and GM (a), their quantum mechanical analog (b), and their classical analog (c). (d) and (e) are plots of Fano shapes for different S values and signs for FWHM of 6nm and zero background reflectivity.

To conclude, a narrowband tunable filter with high contrast and tuning range is proposed based on a resonating planar multilayered structure that exhibits transparency window in reflection with a controllable FWHM. Examples are given showing tunability over wide spectral range (>500nm in the visible and near infrared) using a modulation layer of submicron thickness or small angular scan. The phenomenon is observed in TE and TM polarizations with much higher contrast peak in TE. Analysis confirmed by field distribution calculations show that coupling between waveguide modes and lossy surface electromagnetic waves explains the observed phenomenon followed by Fano resonance analysis. The wide tuning range with high contrast is mainly achieved using an absorptive layer with high ratio between the imaginary to real parts of the dielectric constant that enables excitation of lossy surface waves known to exist over a wide spectral band in thin films. To avoid large losses, it is found that the lossy layer should be ultrathin (6nm Cr layer in TE or 7nm Al layer in TM for example). The tuning is achieved either by small angular scan of less than 2 degrees or by modulating the refractive index or thickness of the submicron thick waveguide layer from the visible till the near infrared range and in principle it can be designed to operate in any spectral range. Such a thin variable index or variable thickness layer can allow tuning at ultrahigh speed using conventional electrooptic, magnetooptic, piezoelectric or thermooptic materials at relatively low external fields. Scanning over such small angular range of 2 degrees can also be achieved in high rate these days using microelectromechanical (MEMS) or liquid crystal scanners. Obtaining such a narrowband fast tunable filter over wide range can boost the field of hyperspectral imaging, Raman imaging, infrared sensing, diagnostics and many other applications. It should also be noted that since many materials have large ratio of the imaginary to real part of the dielectric function in the UV, infrared, THz and microwave regions, this concept should allow designing tunable filters and modulators for these regions as well. It should be mentioned that the intellectual property related to the concept proposed here and its potential applications are protected in a provisional patent application and a PCT [21]. Attempts to manufacture it and demonstrating its functionality are on-going and will be reported shortly.

**Acknowledgements.** I am grateful to Dr. Atef Shalabney for useful discussions and sanity checks and to Mr. Mohammad Abutoama for providing the field calculations using COMSOL.